\documentclass[aps,prl,superscriptaddress,twocolumn,showpacs]{revtex4}
\usepackage{amsmath,graphicx,bm}
\usepackage[colorlinks,linkcolor=blue,anchorcolor=blue,citecolor=red]{hyperref}

\begin{document}
    
\title{Ultrafast Switching of Antiferromagnets via Spin-transfer Torque}
\author{Ran Cheng}
\email{chengr@andrew.cmu.edu}
\affiliation{Department of Physics, Carnegie Mellon University, 5000 Forbes Avenue, Pittsburgh, PA 15213}

\author{Matthew W. Daniels}
\affiliation{Department of Physics, Carnegie Mellon University, 5000 Forbes Avenue, Pittsburgh, PA 15213}
\author{Jian-Gang Zhu}
\affiliation{Department of Electrical and Computer Engineering, Carnegie Mellon University, 5000 Forbes Avenue, Pittsburgh, PA 15213}
\author{Di Xiao}
\email{dixiao@cmu.edu}
\affiliation{Department of Physics, Carnegie Mellon University, 5000 Forbes Avenue, Pittsburgh, PA 15213}


\begin{abstract}
Picosecond switching of the staggered antiferromagnetic order is shown to be realizable through spin-transfer torques from a short current pulse. The coupled dynamics of sublattice magnetization is mapped onto a classical pendulum subject to gravity and a driving pulse, where switching occurs if the pendulum acquires sufficient kinetic energy during the pulse to overcome the maximum of the effective gravity potential. The optimal switching scheme is explored through the dependence of switch angle and magnetic loss on the duration and strength of the current pulse. The physics discussed here provides a general route towards multi-functional THz applications via the spin-transfer torque in antiferromagnetic materials.
\end{abstract}

\maketitle

\textit{\textbf{Introduction}.}---Ultrafast manipulation of magnetic states is a fundamental and lasting issue of magnetic device design. Recently, attention has been increased in ultrafast switching of antiferromagnets (AFs) using a laser pulse, in which the magnetic moments are first excited by a short pulse that lasts hundreds of femtoseconds, then magnetic switching takes place through inertia after the pulse is turned off~\cite{ref:Rasing,ref:NiOswitch,ref:Nowak}. This switching mechanism exploits the giant exchange coupling between neighboring magnetic moments commonly found in AFs. Since the switching occurs on the picosecond time scale, which is orders of magnitude faster than the conventional switching of ferromagnets, it opens a new avenue for ultrafast recording and processing of magnetically stored information using antiferromagnetic materials.

However, to build a viable device using AFs, it is desirable to ask if the laser impetus can be replaced by an electrical current. A promising candidate is the spin-torque magnetoresistive random-access memory (MRAM), where the magnetization is driven by the spin-transfer torque (STT)~\cite{ref:MRAM}. While spin-torque MRAM has only been realized using ferromagnets, recent progress in both experiments~\cite{ref:Tsoi,ref:Urazhdin} and theories~\cite{ref:Arne,ref:Gomonay,ref:MacD,ref:Duine,ref:Ran,ref:Jacob,ref:Manchon,ref:Yaroslav,ref:SPAF} has pointed out the possibility of current-induced excitations of AFs. Based on a microscopic calculation of the electron scattering across a normal metal/AF interface~\cite{ref:SPAF}, it has been demonstrated that a precessing staggered field pumps spin current into the adjacent normal metal, and \textit{vice versa}: a spin accumulation impinging on an AF drives the coherent dynamics of the staggered field. The STTs discovered in AFs not only shed light on the mutual dependence between electron transport and magnetization dynamics, but also opens up the exciting possibility of electric control of AF devices.

In this paper, we show that the staggered field of an AF can be switched within several picoseconds by a short current pulse with spin polarization perpendicular to the easy-plane. The switching process is first investigated by solving the coupled Landau-Lifshitz-Gilbert (LLG) equations, and then further elucidated by an effective pendulum model. The material estimations are based on the widely studied room temperature AF insulator NiO, and the optimal switching is discussed by considering how the switch angle and magnetic loss depend on the pulse duration and STT magnitude.  In addition, a THz nano-oscillator based on the perpendicular geometry is studied. The physics discussed here provides a general route towards multi-functional THz applications via the STT in antiferromagnetic materials.

\begin{figure}[t]
 \centering
 \includegraphics[width=\linewidth]{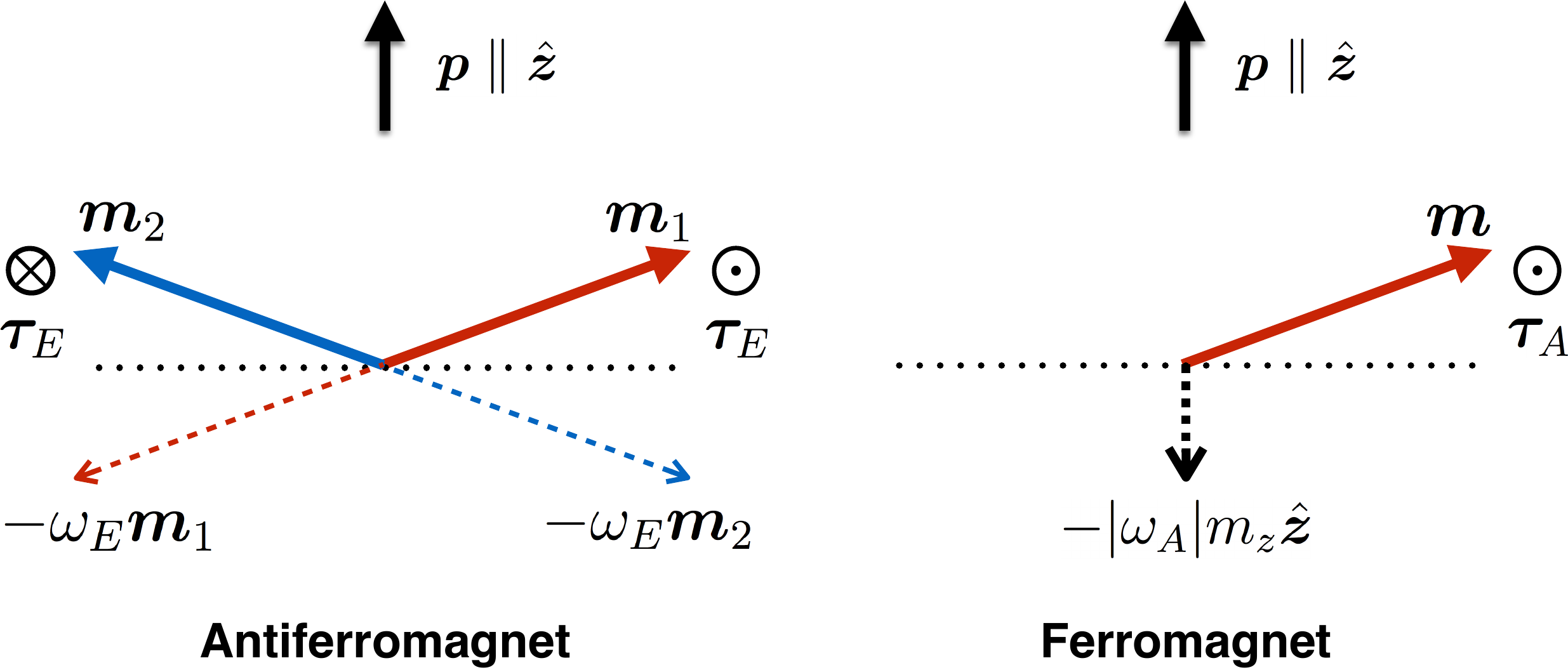}
 \caption{(Color online) For $\bm{p}\parallel\hat{\bm{z}}$, an AF precession is implemented by the exchange torque, whereas a ferromagnetic precession resorts to the demagnetization field.}
 \label{fig:compare}
\end{figure}

\textit{\textbf{Dynamics}.}---For easy-plane AFs such as NiO and MnO, suppose the hard axis is $\bm{\hat{z}}$ and the in-plane easy-axis is $\bm{\hat{x}}$. Scaling everything with frequency, we express the out-of-plane anisotropy by $\omega_A<0$, the in-plane anisotropy by $\omega_a>0$, and the Heisenberg exchange interaction by $\omega_E>0$. Within the macro-spin approximation, the AF is characterized by two classical vectors $\bm{m}_1$ and $\bm{m}_2$ representing the magnetization of the two sublattices. Their dynamics is captured by the coupled LLG equations
\begin{subequations}
 \begin{align}
  \dot{\bm{m}}_1&=\omega_E\bm{m}_1\times\bm{m}_2+\omega_am_{1x}\hat{\bm{x}}\times\bm{m}_1 \notag\\
  &\qquad\qquad+\omega_Am_{1z}\hat{\bm{z}}\times\bm{m}_1+\alpha\bm{m}_1\times\dot{\bm{m}}_1, \label{eq:LLG1}\\
  \dot{\bm{m}}_2&=\omega_E\bm{m}_2\times\bm{m}_1+\omega_am_{2x}\hat{\bm{x}}\times\bm{m}_2 \notag\\
  &\qquad\qquad+\omega_Am_{2z}\hat{\bm{z}}\times\bm{m}_2 +\alpha\bm{m}_2\times\dot{\bm{m}}_2, \label{eq:LLG2}
 \end{align}
\end{subequations}
where $\alpha$ is the phenomenological Gilbert damping constant. We define the staggered field as $\bm{\ell}=(\bm{m}_1-\bm{m}_2)/2$, and the magnetization as $\bm{m}=(\bm{m}_1+\bm{m}_2)/2$. By definition, they are subject to the constraints: $\bm{m}\cdot\bm{\ell}=0$ and $m^2+\ell^2=1$. In the exchange limit, $|\bm{m}|\ll|\bm{\ell}|$ so that $\ell^2\approx1$, and thus $\bm{\ell}\cdot\dot{\bm{\ell}}\approx0$. The recombination of Eqs.~\eqref{eq:LLG1} and~\eqref{eq:LLG2} gives the dynamics of $\bm{m}$ and $\bm{\ell}$
\begin{subequations}
 \begin{align}
  \dot{\bm{m}}&=\omega_a\hat{\bm{x}}\times(m_x\bm{m}+\ell_x\bm{\ell})+\omega_A\hat{\bm{z}}\times(m_z\bm{m}+\ell_z\bm{\ell}) \notag\\
  &\qquad+\alpha\bm{\ell}\times\dot{\bm{\ell}}, \label{eq:mLLG}\\
  \dot{\bm{\ell}}&=2\omega_E\bm{\ell}\times\bm{m}+\omega_a\hat{\bm{x}}\times(m_x\bm{\ell}+\ell_x\bm{m}) \notag\\
  &\qquad+\omega_A\hat{\bm{z}}\times(m_z\bm{\ell}+\ell_z\bm{m}), \label{eq:nLLG}
 \end{align}
\end{subequations}
where higher order damping terms like $\alpha\bm{n}\times\dot{\bm{m}}$, $\alpha\bm{m}\times\dot{\bm{n}}$, and $\alpha\bm{m}\times\dot{\bm{m}}$ have been neglected since $|\bm{m}|\ll|\bm{\ell}|$. In Ref.~\cite{ref:SPAF}, we derived the STTs that exert on $\bm{m}$ and $\bm{\ell}$ in dimensions of frequency as
\begin{subequations}
 \begin{align}
  &\bm{\tau}_m=-\frac{a^3}{e\mathcal{V}}G_r\bm{\ell}\times(\bm{\ell}\times\bm{V}_s), \label{eq:taum}\\
  &\bm{\tau}_{\ell}=-\frac{a^3}{e\mathcal{V}}G_r\bm{\ell}\times(\bm{m}\times\bm{V}_s), \label{eq:taun}
 \end{align}
\end{subequations}
where $G_r$ represents the real part of the spin-mixing conductance, $\mathcal{V}$ is the system volume, and $\bm{V}_s$ is the spin voltage that impinges on the normal metal/AF interface. While Eqs.~\eqref{eq:taum} and~\eqref{eq:taun} are supposed to be added to Eqs.~\eqref{eq:mLLG} and~\eqref{eq:nLLG}, they can equally well be decomposed into $\bm{\tau}_1=\omega_s\bm{m}_1\times(\bm{p}\times\bm{m}_1)$ and $\bm{\tau}_2=\omega_s\bm{m}_2\times(\bm{p}\times\bm{m}_2)$ and added to Eqs.~\eqref{eq:LLG1} and~\eqref{eq:LLG2}, where $\bm{p}$ is the unit vector of the spin polarization, and $\omega_s=a^3G_rV_s/(e\mathcal{V})$ scales linearly with the current density and inversely with the film thickness. When $\bm{p}\parallel\hat{\bm{z}}$, the two torques drag $\bm{m}_1$ and $\bm{m}_2$ slightly out-of-plane as in Fig.~\ref{fig:compare}, so that the exchange interaction generates precessional torque on the magnetic moments. It is this torque that switches the staggered field. Restricted by symmetry, the magnetization $\bm{m}$ develops only an out-of-plane component $\bm{m}=m_z\hat{\bm{z}}$. Correspondingly, the staggered field has only in-plane components so that $\ell_z = 0$. These are confirmed by a straightforward numerical simulation of Eqs.~\ref{eq:LLG1} and~\ref{eq:LLG2} (see Supplementary Materials~\cite{ref:supp}). We mention in passing that no appreciable difference is observed in the switching behavior between a compensated interface where both sublattices are subject to STT and an uncompensated interface where only one of the two sublattice is affected by the STT~\cite{ref:supp}.

\begin{figure}[t]
 \centering
 \includegraphics[width=\linewidth]{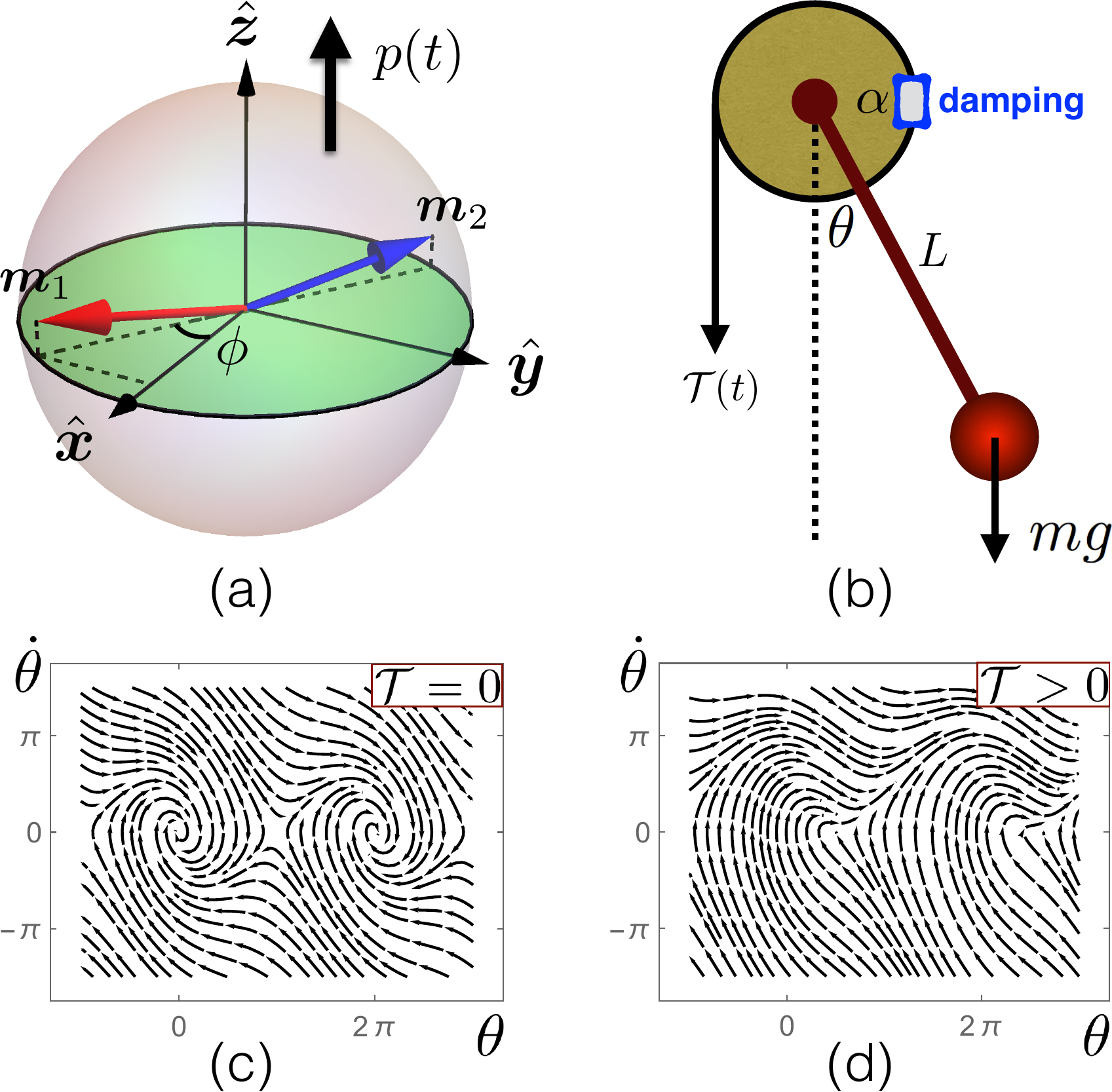}
 \caption{(Color online) Effective model of the staggered field switching. (a) perpendicular STTs cant $\bm{m}_{1,2}$ slightly out of the basal plane. (b) by $\theta=2\phi$, the switching is mapped onto a pendulum subjected to gravity, damping, and a driving torque. (c) and (d): phase portraits of the pendulum for $\mathcal{T}=0$ and $\mathcal{T}>0$ (large enough to open up the attractor at origin), respectively.}
 \label{fig:sinHO}
\end{figure}

The system is now characterized by three variables $(m_z,\ell_x,\ell_y)$. To formulate an effective description of the switching, we eliminate $m_z$ in terms of $\bm{\ell}$. Taking the cross product of $\bm{\ell}$ on Eq.~\eqref{eq:nLLG} leads to
\begin{align}
 \bm{m}=\frac{1}{2\omega_E+|\omega_A|+\omega_a\ell_x^2}\ \dot{\bm{\ell}}\times\bm{\ell}. \label{eq:mexpr}
\end{align}
In typical transition metal oxides such as NiO, $\omega_a$ and $|\omega_A|$ are orders of magnitude smaller than $\omega_E$, thus we can safely ignore the last two terms in the denominator of Eq.~\eqref{eq:mexpr}. Substituting Eq.~\eqref{eq:mexpr} into Eq.~\eqref{eq:mLLG} results in
\begin{align}
 \bm{\ell}\times\left[ \ddot{\bm{\ell}}-\omega_R^2\ell_x\hat{\bm{x}}+2\alpha\omega_E\dot{\bm{\ell}}+2\omega_E\omega_s(\bm{p}\times\bm{\ell}) \right]=0,
 \label{eq:central}
\end{align}
where $\omega_R\equiv\sqrt{2\omega_a\omega_E}$ defines the frequency of in-plane oscillation in an AF~\cite{ref:AFMRNiO}. Equation~\eqref{eq:central} describes the effective dynamics of the staggered field $\bm{\ell}$ in the exchange limit. If we focus only on the in-plane rotation of $\bm{\ell}$, the degrees of freedom are reduced to one and Eq.~\eqref{eq:central} can be further simplified. Denote the in-plane polar angle by $\phi$ so that $(\ell_x,\ell_y)=(\cos\phi,\sin\phi)$; then Eq.~\eqref{eq:central} gives
\begin{align}
 \ddot{\phi}+\frac{\omega_R^2}{2}\sin(2\phi)+2\alpha\omega_E\dot{\phi}=2\omega_E\omega_sp(t), \label{eq:HOorig}
\end{align}
which is a damped non-linear oscillator with a driving force. Under the variable change $\phi\equiv\theta/2$, Eq.~\eqref{eq:HOorig} is recast into
\begin{align}
 \ddot{\theta}+\omega_R^2\sin(\theta)+2\alpha\omega_E\dot{\theta}=4\omega_E\omega_sp(t), \label{eq:HO}
\end{align}
which is reminiscent of a pendulum subject to gravity and driving force as depicted in Fig.~\ref{fig:sinHO} (b). The pendulum is connected to a fixed pulley through a rigid but massless rod. If we twirl the pulley quickly by a short pulse, the motion of the system is described by $J\ddot{\theta}+mgL\sin\theta+\alpha\dot{\theta}=\mathcal{T}(t)$, where $J$ is the moment of inertia, $mg$ is the gravity of the ball, $L$ is the length of the rod, $\alpha$ is the damping constant, and $\mathcal{T}(t)$ is the driving torque exerting on the pendulum. Upon the analogies: $\omega_E\rightarrow1/(2J)$, $\omega_a\rightarrow mgL$, and $\omega_sp(t)\rightarrow\mathcal{T}(t)/2$, Eq.~\eqref{eq:HO} is mapped exactly onto the pendulum motion, by which the staggered field switching ($\pi$ rotation of $\phi$) is represented by the crossing of the gravity maximum by the pendulum ($2\pi$ rotation of $\theta$). The phase portrait of the pendulum with (without) the driving torque $\mathcal{T}$ is plotted in Fig.~\ref{fig:sinHO} 2(d) [Fig. 2(c)]. When $\mathcal{T}=0$, the phase point $(0, 0)$ is an attractor; when $\mathcal{T}$ is sufficiently large, the attractor is broken up and flows towards the right, which enables the switching.

During the pulse, energy is transferred from conduction electrons to the sublattice magnetization $\bm{m}_1$ and $\bm{m}_2$ via STT, and then stored in the exchange energy as they are canted non-collinearly. It is the releasing of this stored energy that provides an effective inertia to the motion afterwards, through which the system surmounts the anisotropy barrier and finally relaxes to a new configuration. On the other hand, as shown in Fig.~\ref{fig:compare}, a ferromagnetic switching under perpendicular spin polarization is engendered by the demagnetization field, which is orders of magnitude slower. While the exchange mechanism is intrinsic to AFs, the demagnetization field is sensitive to the shape anisotropy of ferromagnetic films.

\begin{figure}[t]
 \centering
 \includegraphics[width=0.9\linewidth]{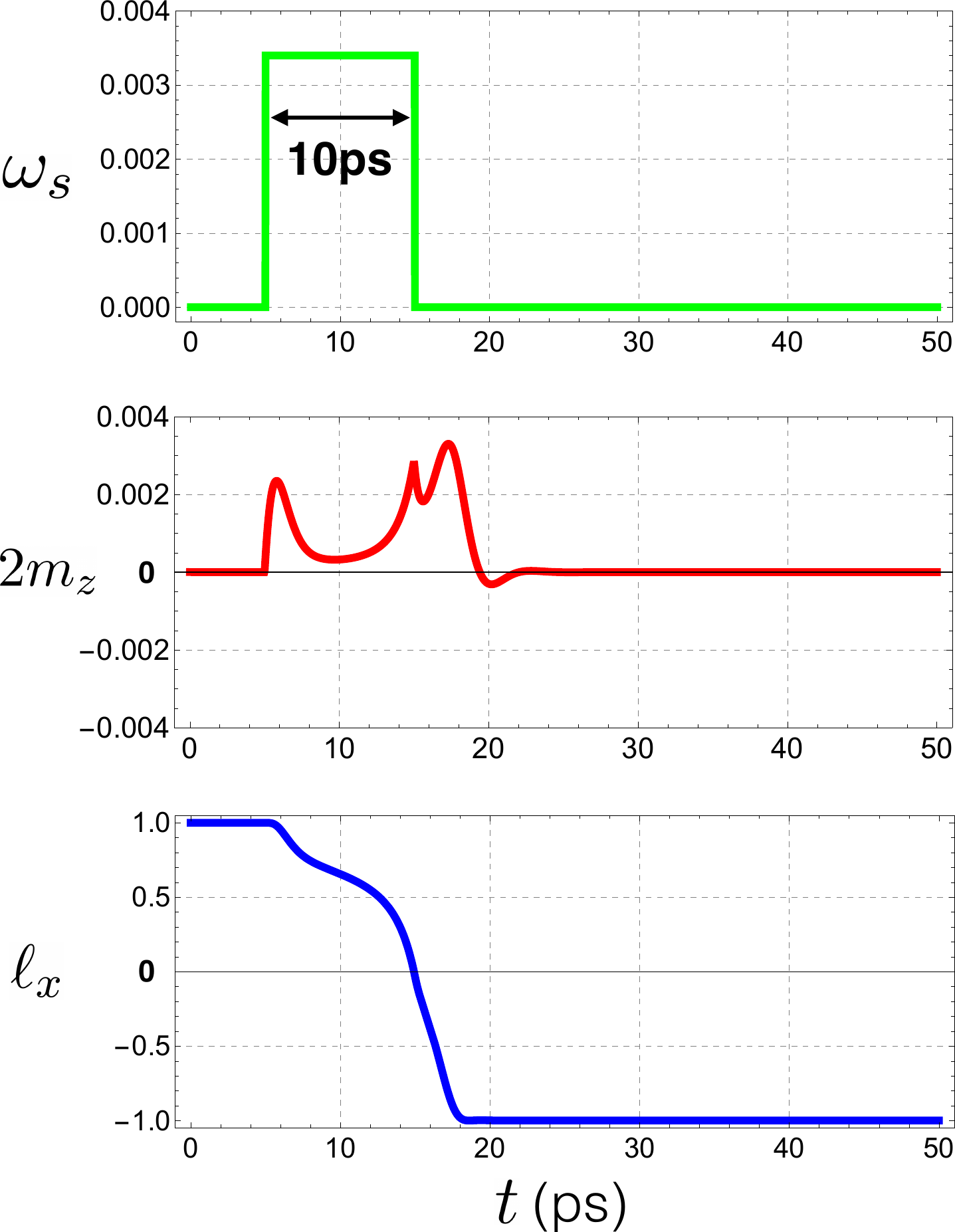}
 \caption{(Color online) Out-of-plane magnetization ($2m_z$) and staggered field projected on the easy-axis as functions of time in picoseconds. Pulse duration $T=10$ps; STT strength $\omega_s=0.0034$; Gilbert damping $\alpha=0.005$. Parameters for NiO: $\omega_E/(2\pi)=27.4$THz, $\omega_a/(2\pi)=1$GHz.
 }
 \label{fig:timeevolve}
\end{figure}

\textit{\textbf{Switching}.}---With the effective pendulum model, we are able to perform a quantitative analysis of the staggered field switching. Using material parameters from NiO~\cite{ref:AFMR,ref:AFMRNiO}, assuming $\alpha=0.005$~\cite{comment}, we plot in Fig.~\ref{fig:timeevolve} the time evolutions of the out-of-plane magnetization $2m_z(t)$ and the $x$-component of the staggered field $\ell_x(t)$. The switching process is characterized by the second of these, which is composed of two steps as follows: (1) the pulse drives the two magnetic moments slightly out-of-plane, thus they rotate under the exchange torque. (2) the staggered order moves to the opposite direction due to the inertia accumulated in the first step. The total energy pumped into the system via STT finally dissipates away through the Gilbert damping.

In Fig.~\ref{fig:timeevolve}, the canted magnetization during the switching process is less than 1\%, such that the approximation $\ell^2\approx1$ is well respected. For small damping $\alpha=0.002$ (not plotted), the flip is followed by a ringing tail; for $\alpha=0.005$ plotted in Fig.~\ref{fig:timeevolve}, the ringing effect is suppressed and the pattern resembles critical damping. Note that for small oscillations around the easy-axis, the critical damping for NiO is roughly $\alpha^{crit.}=0.0085$; but for large angle rotations here, we observe a critical behavior for $\alpha$ smaller than $\alpha^{crit.}$.

NiO has rocksalt structure, and all magnetic moments are attributed to Ni atoms that exhibit fcc configuration. The \{111\}-planes are ferromagnetically ordered and stack in an alternative manner, forming a layered AF. The easy spin direction is $\langle11\bar{2}\rangle$. To prepare the accumulation of conduction electron spins, we may adopt the spin Hall effect in a heavy metal~\cite{ref:SHdevice,ref:SHNO}. This method creates an in-plane spin polarization, thus the NiO needs to be grown in the $\{11\bar{2}\}$-direction to satisfy the required geometry in Fig.~\ref{fig:compare} and Fig.~\ref{fig:sinHO} (a). A much more effective spin accumulation can be achieved by driving the surface states of a topological insulator~\cite{ref:TIdevice}, which will significantly reduce the required current density. But in the following, numerics are restricted to the spin-Hall-driven NiO switching on a Pt/NiO interface.

Attention should be paid to the proper choice of the STT strength. The staggered field will retrieve to the initial position if the STT is insufficient, and will overshoot if it is too strong. For a Pt(20nm)/NiO(3nm) heterostructure assuming no disorder and roughness on the interface~\cite{ref:SPAF}, the $\omega_s$ in Fig.~\ref{fig:timeevolve} is converted to a current density of $6\sim7\times10^7$A/cm$^2$. The required STT for a proper switching becomes smaller when the pulse duration becomes longer. But for very long pulses, the required STT ceases to reduce further. Therefore, to better understand the pulse dependence of the switching, we resort to the terminal angle of the staggered field and the total magnetic loss due to Gilbert damping as functions of the pulse duration and the strength of STT.

\begin{figure}
 \centering
 \includegraphics[width=\linewidth]{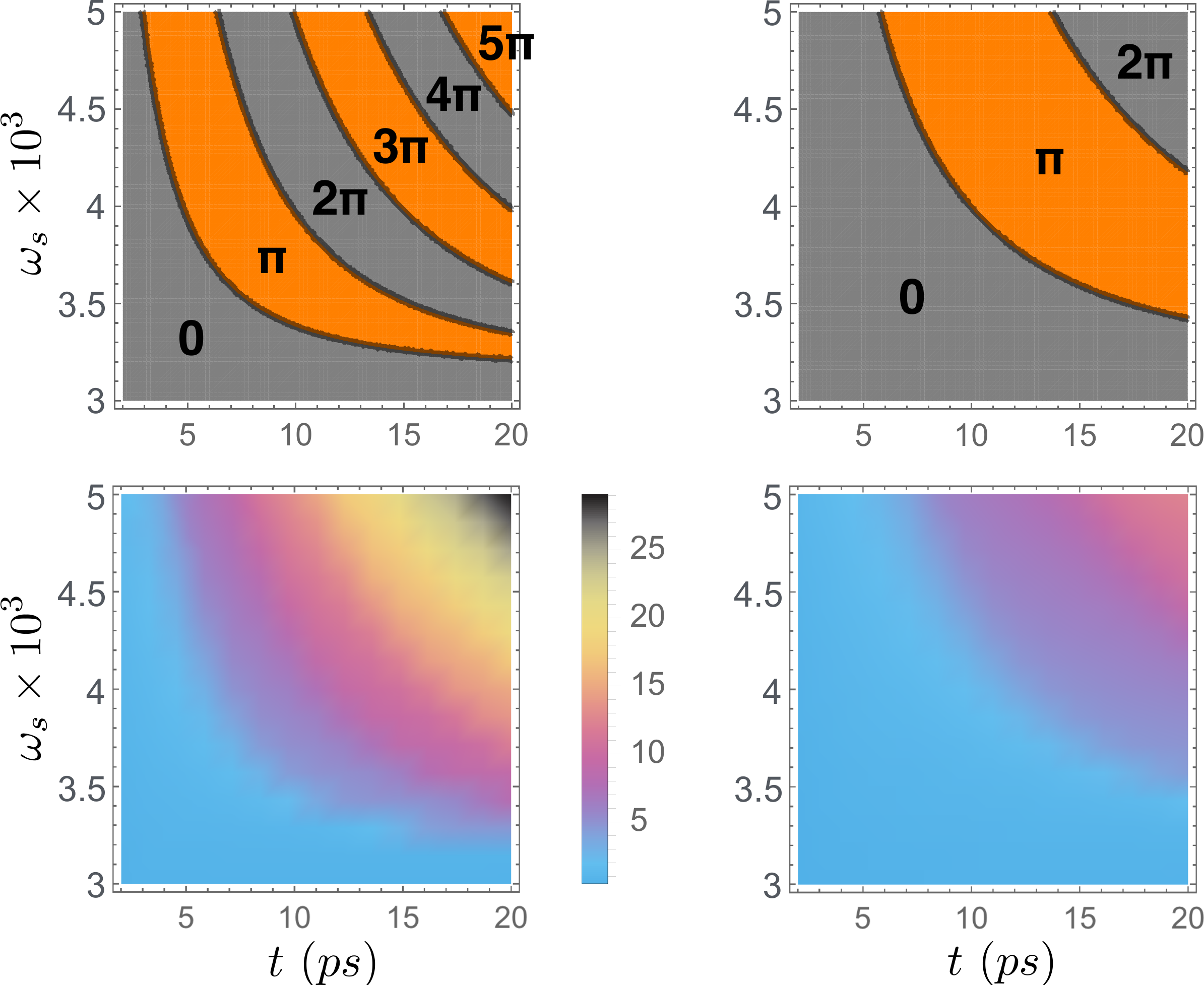}
 \caption{(Color online) Upper left (right): terminal angle of $\bm{\ell}$ as a function of the pulse duration from 2ps to 20ps, and $\omega_s$ from 0.003 to 0.005 for $\alpha=0.005$ ($\alpha=0.01$). Lower left (right): total magnetic loss due to Gilbert damping for $\alpha=0.005$ ($\alpha=0.01$).
 }
 \label{fig:Flip}
\end{figure}

As shown by Fig.~\ref{fig:Flip}, a smaller damping gives rise to a narrower window of the desired rotation (lowest orange region marked by $\pi$). In real experiments, amplitude fluctuation of an electric pulse is inevitable, thus to stabilize the functioning of the device, extremely small damping is not favorable since it may easily lead to overshoot. The total magnetic loss has a similar pattern as the terminal angle, hence it can hardly provide a practically independent criterion for the optimal STT. However, if Joule heating in the normal metal is taken into account, which scales as $\omega_s^2T$ ($T$ is the pulse duration), we obtain a different pattern compared to Fig.~\ref{fig:Flip}. To minimize Joule heating with given magnetic loss, a shorter pulse with relatively stronger current density is preferred. Hopefully, with progress in nano-technology, picosecond or shorter current pulses can be realized in the future. In that case, the staggered field acquires a sufficiently large kinetic energy (angular velocity) long before it reaches the potential maximum of the in-plane anisotropy. The process would then be similar to the laser-pulse-induced AF switching~\cite{ref:Rasing,ref:NiOswitch,ref:Nowak}. However, a laser pulse couples to the AF via the weak Zeeman interaction, while STT rests on the strong exchange coupling between conduction electrons and local moments.

\textit{\textbf{Terahertz Oscillator}.}---Another possible application associated with the perpendicular geometry illustrated by Fig.~\ref{fig:compare} is the terahertz nano-oscillator. If we replace the pulse current by a dc current in the Pt, the NiO will undergo a continual rotation. In Fig.~\ref{fig:sinHO} (b), it amounts to a continual rotation of the pendulum against gravity under a constant driving torque. Since $\omega_a\ll|\omega_A|$ in NiO, the in-plane anisotropy barrier causes only tiny non-uniformity in the angular velocity, which we can simply ignore. A continual rotation occurs when STT exactly compensates Gilbert damping. By a simple geometry, the compensation condition is $\omega_s=2\alpha\omega_E\sin\vartheta$, where $\vartheta$ is the azimuthal angle of $\bm{\ell}$ with respect to the hard $\hat{\bm{z}}$-axis. To leading order, the achieved oscillation frequency is linear in the STT (or the applied current):
\begin{equation}
\omega_{\text{rot.}}=\omega_s/\alpha \;, \label{eq:rot}
\end{equation}
which is independent of the exchange interaction. For $\alpha=0.002$ and a Pt(20nm)/NiO(3nm) structure discussed above, a 1THz oscillation requires a current density of roughly $2\sim3\times10^8$A/cm$^2$; a thinner NiO and/or heavy metal with larger spin Hall angle will scale down the required current density. Moreover, the linearity of Eq.~\eqref{eq:rot} remains robust up to about 10THz, bearing an extraordinarily large range of linear control on the nano-oscillator.

Just as with the nano-oscillator predicted by the authors of Ref.~\cite{ref:SPAF}, the oscillation direction is determined by the current direction (a binary selection rule), which is unattainable in ferromagnetic materials. However, the excited modes are fundamentally different here. In Ref.~\cite{ref:SPAF}, the spin accumulation in the normal metal is parallel to the staggered field, which excites the AF resonance eigenmodes. In this paper, the spin accumulation is perpendicular to the staggered field, and the switching is realized by exciting the spin superfluid mode~\cite{ref:Yaroslav}.

\textit{Acknowledgments}---This work is partially supported by the U.S. Department of Energy, DE-SC0012509 (D.X.) and the National Science Foundation, Office of Emerging Frontiers in Research and Innovation, EFRI-1433496 (M.W.D.). R.C. is grateful to J. Zhou and Y. You for helpful discussions.


\begin{thebibliography}{20}
\bibitem{comment} In the presence of Gilbert damping, the renormalized frequency of the in-plane mode is $\sqrt{2\omega_a\omega_E-\alpha^2\omega_E^2}$. The bare frequency $\sqrt{2\omega_a\omega_E}/(2\pi)$ is $234$GHz, while the observed frequency in Ref.~\cite{ref:NiOswitch} is $140$GHz, thus $\alpha$ is $0.0068$. The actual $\alpha$ varies from sample to sample, but we estimate it to be in the range $0.005\sim0.01$.
\end{thebibliography}
\end{document}